# The Challenges of Deploying Artificial Intelligence Models in a Rapidly Evolving Pandemic

Yipeng Hu[1,2,4] Joseph Jacob[1,3] Geoffrey JM Parker[1,5,6] David J Hawkes[1,2,4] John R Hurst[3] Danail Stoyanov[1,2,5]

[1] UCL Centre for Medical Image Computing,
[2] Wellcome / EPSRC Centre for Interventional and Surgical Sciences,
[3] UCL Respiratory,
[4] Department of Medical Physics and Biomedical Engineering, and
[5] Department of Computer Science,
University College London, Gower Street, London WC1E 6BT, United Kingdom
[6] Bioxydyn Limited, Pencroft Way, Manchester, M15 6SZ, United Kingdom

Correspondence: yipeng.hu@ucl.ac.uk

The COVID-19 pandemic, caused by the severe acute respiratory syndrome coronavirus 2, emerged into a world being rapidly transformed by artificial intelligence (AI) based on big data, computational power and neural networks. The gaze of these networks has in recent years turned increasingly towards applications in healthcare. It was perhaps inevitable that COVID-19, a global disease propagating health and economic devastation, should capture the attention and resources of the world's computer scientists in academia and industry. The potential for AI to support the response to the pandemic has been proposed across a wide range of clinical and societal challenges [1], including disease forecasting, surveillance and antiviral drug discovery. This is likely to continue as the impact of the pandemic unfolds on the world's people, industries and economy but a surprising observation on the current pandemic has been the limited impact AI has had to date in the management of COVID-19. This correspondence focuses on exploring potential reasons behind the lack of successful adoption of AI models developed for COVID-19 diagnosis and prognosis, in front-line healthcare services. We highlight the moving clinical needs that models have had to address at different stages of the epidemic, and explain the importance of translating models to reflect local healthcare environments. We argue that both basic and applied research are essential to accelerate the potential of AI models, and this is particularly so during a rapidly evolving pandemic. This perspective on the response to COVID-19, may provide a glimpse into how the global scientific community should react to combat future disease outbreaks more effectively.

# The evolving clinical need

The clinical management of COVID-19 has spanned various stages including anticipation, early detection, containment, mitigation, together aiming towards eventual eradication [2]. Each stage differs in its measured and actual disease prevalence, which directly impacts the availability of clinical resources, and over a matter of weeks, clinical priorities can fluctuate rapidly. Priorities may range from the provision of robust diagnoses, maintenance of infection control to ensuring availability of facilities for mechanical ventilation. Such rapid changes occuring in tandem with enhanced knowledge of virus behaviour and availability of supporting data, have meant that the outputs required of predictive AI models need to constantly evolve. Accordingly, the AI models that are most urgently needed and which can be feasibly built are likely to be different at each epidemic stage.

**The anticipation and early detection stages:** With a relatively low number of positive cases and many potentially asymptomatic cases during the early stages of the pandemic, a highly sensitive diagnostic AI model to detect COVID-19 would have been useful. The lack of pre-existing data from this new disease, meant the feasibility of building such new AI models to determine diagnosis or prognosis remains an open challenge, yet one that is of particular interest to an AI community focused on breaking the existing barriers between data domains using machine learning. Generalising AI models to unseen data (inference), to data coming from different distributions (domain adaptation, transfer learning) and to data with limited or no labels (semi- or unsupervised learning), e.g. [3], are all priority areas in the technical development of AI and such non-fully-supervised methods may benefit all the scenarios discussed. These early stages of a new disease have been described as overlooked periods in the general management of infectious diseases [4]. The AI community should design strategies and methodologies for rapid deployment in the event of future epidemic threats, to enable data collection, model training, testing and wide deployment to be as efficient as possible.

**The containment and mitigation stages:** During the containment and mitigation of COVID-19, data have become increasingly available with the exponential growth of confirmed positive cases. During this time it is essential to rapidly curate sizable training data sets and develop stable, well-performing AI models that can respond to emerging clinically urgent needs, such as rapid, consistent patient triage at scale across a health service.

Reverse transcription polymerase chain reaction (RT-PCR) tests via nasopharyngeal swabbing with nearly 100% specificity are considered the diagnostic ground-truth for COVID-19. However RT-PCR has limited negative predictive value with variable availability and diagnostic speed. Alternative methodologies for diagnosing COVID-19 include medical imaging techniques such as computed tomography (CT) and chest radiographs (CXR) [5]. Some groups have also explored point-of-care ultrasound albeit with limitations. Driven by data availability, the focus of AI work in COVID-19 has centered on RT-PCR-labelled diagnostic models, e.g. [6], or the automated evaluation of clinical/imaging features, e.g. lung involvement on CT imaging [7]. Those developing AI-assisted diagnostic tools must recognise that very high diagnostic accuracies are required to demonstrate added value above and beyond existing clinical imaging and RT-PCR tests.

It is also important to question which prognostic outcomes require greatest prioritisation during this period. The majority of existing AI models aim to predict hospitalisation and mortality, e.g. [8], using predictors such as age, gender, blood biomarkers, pre-existing co-morbidities and imaging [5]. In resource constrained clinical environments, there is great value in predicting resource consumption as a "surrogate" prognostic outcome, as a lack of personal protective equipment (PPE) for example can directly affect community prognoses [9]. Intuitive candidate outcome measures for AI models might include time spent on mechanical ventilators and/or within intensive care units. Yet as knowledge of COVID-19 has grown, early intubation of a patient has diminished in priority in the care pathway. Similarly, with limitations in resources pragmatic choices have had to be made regarding patient selection for intensive care unit admission. Evolving management strategies such as these have a real-time impact on the outcomes that AI models aim to predict. Disease progression (or regression following treatment) models can be trained using time series data, such as longitudinal CT images [10], to quantify the likelihood of development of severe pneumonia and acute myocardial injury, two leading causes of mortality [11], and the cytokine release syndrome. A lesson from the COVID-19 pandemic has been that AI models motivated merely by the practical convenience of utilising acquiring labelled data have had limited clinical value.

**The eradication stage and beyond:** At the later eradication stage, constraints in data availability, development time and clinical resources would gradually be eased. The number of positive cases can drop quickly. Yet the need for a real-time, convenient, highly-sensitive screening tool may persist to control transmission and judiciously recognise potential outbreaks.

The requirements for prognostic tools may shift to the identification of patients at risk of developing long term sequelae such as pulmonary fibrosis. Indeed cardiopulmonary, neurological [12] and urological [13] damage are all being recognised following COVID-19 infection. Given the potentially significant health service resource requirements that may result from long-term complications across large swathes of the population, the post-acute phase of COVID-19 will be a critical clinical research area, where AI models may play a central role.

## Translating AI models

A typical AI translation workflow includes model development, model deployment and model adaptation (or model update). The COVID-19 AI research efforts have been concentrated primarily on new model development and the urgency brought about by the pandemic must not override the stringent requirements for clinical deployment [14]. Despite perceived time pressures, rigorous validation is key to ensuring that safety and efficacy are tested; models must be validated before initial deployment, continuously monitored and adapted, when implemented in local healthcare environments and as outcome likelihoods change due to evolving patient management strategies. Failing to adhere to such practice will impede translation and compromise the impact of AI on clinical needs.

**Pre-deployment validation:** Recent COVID-19 AI models have been criticised for a lack of transparency in development and a high likelihood of bias towards nonrepresentative patient populations [5]. Data limitations in availability and quality can be the inherent cause of problems, for example validation data sets with unrealistically high numbers of control cases acquired during the beginning or extremely low numbers of control cases at the peak of the outbreak. These models are unlikely to be directly useful in all stages of the pandemic due to potential bias.

Best practices in rigorous design and analysis of experiments should be adopted for AI model validation. In addition, model interpretation methods help to explain the reasoning of the predictions, e.g. [15, 16], and may also indicate when certain data-driven methods are unlikely to generalise [17]. Model transparency could also be key to addressing regulatory and ethical issues [18, 19].

***Local adaptation:*** It is not uncommon to find that an AI model trained with data from one healthcare centre, or even from multiple centres, does not generalise as well at a new centre. For example, the accuracy of chest x-ray detection represented by the area under the receiver operating characteristic curve, was significantly reduced from 0.93 in a multi-centre internal validation, to 0.82, on external validation data [17]. The pre-deployment external validation reduces the risk of this overfitting problem based on the assumption that the external data represent new local data. However, the local data are likely to have unique characteristics due to centre-specific acquisition features, equipment and protocols, all of which may have differing clinical constraints and requirements - the local healthcare environment [20]. Moreover, temporal differences in data may increase, adversely affecting model accuracy, as the demographic and immunity landscape and clinical practice shift between different stages of the epidemic [21]. AI models therefore should have a continuous monitoring and adaptation strategy to these changing data to maintain their predictive accuracy.

Most proposed AI approaches for COVID-19 diagnosis/prognosis have so far been "locked" algorithms that do not facilitate future adaptation. Model adapting methods from other medical applications should be tested and integrated in these developments, such as transfer learning [22] and model retraining with a small local data set. Recently, the FDA has proposed a new approach to allow AI-based software to adapt and improve from real-world use [23], paving the regulatory pathway to address these local adaptation needs.

# Conclusion

The COVID-19 pandemic has presented numerous challenges to virtually every section of society in all geographic locations. AI can be an enabling technology to support urgent clinical needs in disease diagnosis and prognosis but is reliant on appropriate infrastructure, data management and translational pathways. New international cross-disciplinary collaborations, carefully identifying time-course-and-region-dependent clinical actions in response to COVID-19 can benefit from scientifically sound AI model development, validation and deployment to support local healthcare providers. Safe and responsible translation is the only way for realising the promise for AI models to contribute in combating the current coronavirus pandemic, its aftermath and, potential future clustered outbreaks or comparable healthcare emergencies.

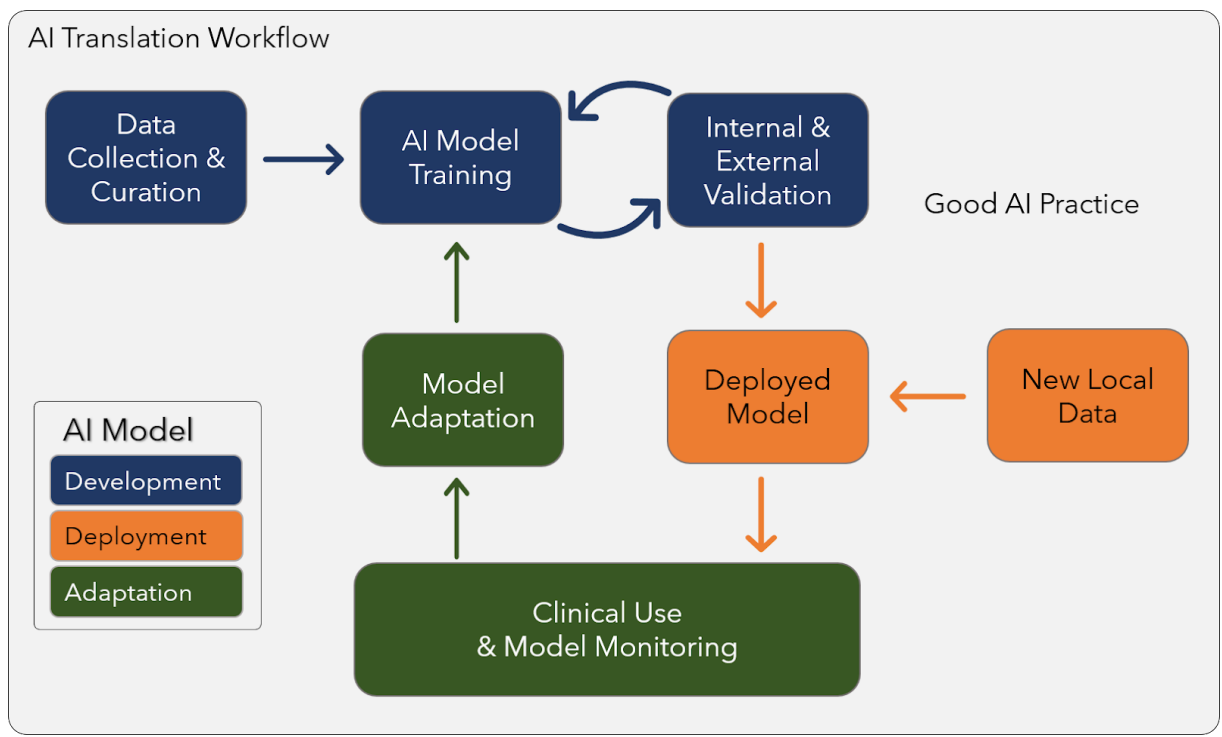

**Competing interest statement**
The authors declare no competing interests in relation to the submitted work. Outside of this submission, J.J. reports consultancy fees from Boehringer Ingelheim and Roche.

**Acknowledgement**
This work is supported by the Wellcome/EPSRC Centre for Interventional and Surgical Sciences (203145Z/16/Z). J.J. was supported by a Wellcome Trust Clinical Research Career Development Fellowship (209553/Z/17/Z) and acknowledges support from the NIHR Biomedical Research Centre at University College London.